\documentclass[]{spie}
\usepackage{cite}

\usepackage{tikz}
\usetikzlibrary{shapes,arrows,dsp,fit,positioning}
\usepackage[americanresistors,americaninductors]{circuitikz}

\dspdeclareoperator{dspshapesine}{
	\draw [line cap=round] (-3/32,0)
		sin (-3/64,1/16) cos (0,0) sin (3/64,-1/16) cos (3/32,0);
	\pgfusepathqstroke
}
\tikzset{dspsine/.style={shape=dspshapesine,line cap=rect,line join=rect,
	line width=\dspblocklinewidth,minimum size=\dspoperatordiameter}}

\usepackage{float} % for [H]
\usepackage{amssymb} % for mathbb
\usepackage{cleveref}
\usepackage{siunitx}

\begin{document}

\title{Digital Active Nulling for Frequency-Multiplexed Bolometer Readout: Performance and Latency}
\author[a]{Graeme Smecher}
\affil[a]{Three-Speed Logic, Victoria, Canada}
\authorinfo{E-mail: gsmecher@threespeedlogic.com}

\author[b]{Tijmen de Haan}
\affil[b]{High Energy Accelerator Research Organization (KEK), Tsukuba, Japan}

\author[c]{Matt Dobbs}
\author[d]{Joshua Montgomery}
\affil[c,d]{McGill University, Qu\'ebec, Canada}

\maketitle

\begin{abstract}
	We consider the stability and performance of a discrete-time
	control loop used as a dynamic nuller in the presence of a relatively
	large time delay in its feedback path.

	Controllers of this form occur in mm-wave telescopes using
	frequency-multiplexed Transition Edge Sensor (TES) bolometers.  In this
	application, negative feedback is needed to linearize a Superconducting
	Quantum Interference Device (SQUID) used as an amplifier. $M$ such
	feedback loops are frequency-multiplexed through a SQUID at distinct
	narrowband frequencies in the MHz region. Loop latencies stem from the
	use of polyphase filter bank (PFB) up- and down-converters and have
	grown significantly as the detector count in these experiments
	increases.

	As expected, latency places constraints on the overall gain $K$ for
	which the loop is stable.  However, latency also creates spectral peaks
	at stable gains in the spectral response of the closed loop.  Near
	these peaks, the feedback loop amplifies (rather than suppresses) input
	signals at its summing junction, rendering it unsuitable for nulling
	over a range of stable gains.
	
	We establish a critical gain $K_C$ above which this amplifying or
	``anti-nulling'' behaviour emerges, and find that $K_C$ is
	approximately a factor of 3.8 below the gain at which the system
	becomes unstable.

	Finally, we describe an alteration to the loop tuning algorithm that
	selects an appropriate (stable, effective for nulling) loop gain
	without sensitivity to variations in analog gains due to component
	tolerances.
\end{abstract}

\keywords{Negative Feedback, Delays, Closed-Loop Systems, Stability Analysis}

\section{Introduction}

At the South Pole Telescope in Antarctica~\cite{bender_year_2018}, and in many
other mm-wave telescopes~\cite{vanderlinde10, Schwan2011, polarbear2014a, Hattori2013,
bleem12a, Abitbol2018}, frequency-multiplexed transition edge sensor (TES)
bolometers convert incident radiation power into an electrical signal for
measurement. TES bolometers achieve extremely high sensitivity by operating
near their superconducting transition temperatures, where their resistance
varies strongly with temperature (and hence, absorbed
radiation)~\cite{Irwin1995}.

To further amplify faint astrophysical signals for digitization by
room-temperature electronics, these telescopes use cryogenic superconducting
quantum interference devices (SQUIDs).  SQUIDs are magnetometers that, when
coupled to an input coil, form exquisitely sensitive transimpedance amplifiers.
However, SQUIDs in this configuration exhibit a nonlinear (approximately
sinusoidal) response to input current. To linearize SQUIDs, their input signals
are typically DC biased with a limited dynamic range to avoid distortion.

In a frequency-multiplexed readout~\cite{Lanting2005, Smecher2012}, each SQUID
amplifies signals from $M$ bolometers at distinct carrier frequencies in the
MHz range.  Designs supporting values of $M$ between 8 and 128 have been
fielded, and future mm-wave telescopes are expected to continue aggressive
growth of the multiplexing factor $M$.

In Ref.~\citenum{dehaan12}, we introduced Digital Active Nulling (DAN), a mechanism for
adaptively nulling the input to a cryogenic SQUID amplifier. DAN linearizes a
SQUID in the presence of multiple narrow-band modulated carriers by dynamically
controlling the amplitude of a nulling tone associated with each carrier. DAN
also reduces the effective input impedance of the SQUID inductor in the bands
surrounding each detector, reducing the impact of parasitic circuit elements in
the cold electronics that can compromise detector stability and response
linearity. A DAN system forms multiple frequency-multiplexed feedback loops
with a single cryogenic summing junction at the SQUID input coil.  Although
each DAN loop operates at RF, its negative feedback operates at baseband in a
small frequency range around each carrier, and the loop bandwidth is as
small as several kHz.

DAN has been successfully deployed on several mm-wave telescopes since its
introduction~\cite{vanderlinde10, polarbear2014a, tomaru12, bleem12a,
Abitbol2018}, and though the means of implementing this signal path has
evolved~\cite{bender_year_2018}, the underlying nulling algorithm has not.

This work investigates DAN dynamic performance and loop stability, focusing
primarily on the impact of latency (a design parameter driven higher by the
efficient design of high-channel-count up- and down-converters) and loop gain
(a tunable parameter that determines nulling efficiency and spectral flatness
across the bolometer bandwidth). As the detector count of mm-wave telescopes
continues to increase, the relationship between these performance
characteristics and the maximum attainable $M$ is increasingly important.

This paper is organized as follows:
\begin{itemize}
	\item
		We describe a frequency-multiplexed bolometer readout system
		operating in DAN mode, to provide both a theoretical foundation
		and a motivating example for the remainder;
	\item
		We reduce the system to a simplified model of the DAN feedback
		loop at baseband, and investigate the system's stability under
		varying latency and loop gain;
	\item
		We investigate the system's suitability as a nulling loop for
		SQUID amplifiers, and impose further loop gain constraints to
		prevent unwanted amplification of signals at the input to the
		SQUID; finally,
	\item
		We describe alterations to the tuning algorithm that allow nearly
		optimal selection of the loop gain, even when the analog gains
		are only approximately known (e.g. perturbations due to component
		variations).
\end{itemize}

\section{Frequency-Multiplexed Bolometer Readout}

\subsection{Cryogenic Electronics}

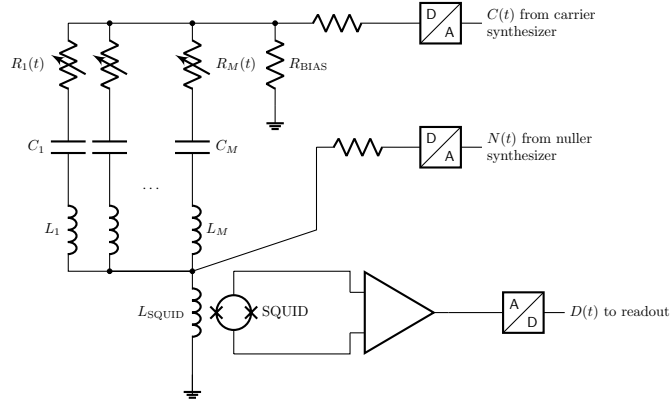
\begin{figure}
	\centering
	\begin{circuitikz}[scale=0.55, transform shape]
	\draw
		% Carrier DAC
		(1,9) -- (6,9) to [R] (9,9) to [dac] (11,9)

		% First resistor
		(1,9) to [vR, -, l_=$R_1(t)$] (1,7)
		to [C, l_=$C_1$] (1,5)
		to [L, l_=$L_1$] (1,3) -- (2,3)
		-- (3,3) -- (4,3)

		% Second resistor
		(2,9) to [vR, *-] (2,7)
		to [C] (2,5)
		to [L, -*] (2,3) -- (4,3)

		% Ellipsis
		(3,5) node[] {$\cdots$}

		% i'th resistor
		(4,9) to [vR, *-, l=$R_M(t)$] (4,7)
		to [C, l=$C_M$] (4,5)
		to [L, l=$L_M$, -*] (4,3)

		(6,7) node[ground] {}
		to [R, -*, l_=$R_\mathrm{BIAS}$] (6,9)

		% SQUID coil
		(4,3) to [L, l_=$L_\mathrm{SQUID}$] (4,1)
		to node[ground] {} (4,0)

		% Nuller
		(7,6) to [R] (9,6) to [dac] (11,6)
		(7,6) |- (7,4) -- (4,3)

		% Readout
		(9,2) node[plain amp] (amp) {}

		(amp.-) |- (5,3)
		(amp.+) |- (5,1)

		(5,3) to [squid,l={~SQUID}] (5,1)

		(amp.out) -- (11,2) to [adc] (13,2)
		;

		% Labels
		\node[right, text width=3cm] at (11,9) {$C(t)$ from carrier synthesizer};
		\node[right, text width=3cm] at (11,6) {$N(t)$ from nuller synthesizer};
		\node[right, text width=3.5cm] at (13,2) {$D(t)$ to readout};

	\end{circuitikz}
	\caption{A simplified schematic of the analog portion of a
		voltage-biased, frequency-multiplexed bolometer readout system.
		A carrier D/A converter (top right) delivers a comb of
		sinusoids $C(t)$ to a bias resistance $R_\mathrm{BIAS} \ll
		R_m(t), 1 \leq m \leq M$, establishing a voltage bias across
		$M$ RLC networks.  In these networks, the time-varying
		resistances $R_m(t)$ are electrical models of the bolometers.
		Each LC filter selects a single frequency from the carrier
		comb, maintaining a distinct bias carrier across each
		bolometer. This bias holds the bolometer in electrothermal
		feedback~\cite{Irwin1995}.  The SQUID provides a measurement of
		the summed current through all $M$ bolometers, which is
		cancelled by a dedicated nuller D/A converter (centre right).
		The residual signal is amplified and digitized for readout.}
	\label{fig:cold-schematic}
\end{figure}

The cryogenic electronics associated with a frequency-multiplexed bolometer
readout system are shown in~\Cref{fig:cold-schematic}.  Incident
millimeter-wave energy is absorbed by an array of $M$ focal-plane bolometers
associated with a single SQUID, each of which acts as a time-varying resistance
$R_1(t), R_2(t), \ldots R_M(t) \approx \SI{1}{\ohm}$. Typical multiplexing
factors range between $M=16$ and $M=128$. (The design of the cryogenic
electronics evolves semi-independently from the warm readout electronics; see
e.g. Ref.~\citenum{Lowitz2018}.)

Each bolometer is paired with a tuned LC circuit, associating that bolometer
with a single carrier frequency within a comb $C(t)$ of synthesized carriers.
As each bolometer's resistance changes, the associated carrier tone is
amplitude modulated to produce sidebands that encode the incident sky signal.

The modulated carrier tones are summed through an inductor $L_\mathrm{SQUID}$.
The resulting magnetic field is converted to a voltage signal by the SQUID,
then amplified and digitized in room-temperature electronics.

The time scale of varying bolometer resistances (hence, the bandwidth of their
associated sidebands) is determined by the dynamic characteristics of the
incident mm-wave signal, which depends on the sky itself, noise contamination
from thermal, optical, or electric sources, the telescope's scanning speed, and
the impact of any modulating elements such as a rotating half-wave plate used
in the system. This time scale is fundamentally limited by the bolometers'
thermal time constants, which are typically in the millisecond range.
Generally, the science signals of interest are captured in several tens of Hz
surrounding each carrier.

To reduce the dynamic range seen by the SQUID and associated electronics, an
auxiliary DAC cancels or ``nulls'' bolometer current at the SQUID using the
signal $N(t)$ (see~\Cref{fig:cold-schematic}).  It is the primary task of the
DAN loop to generate this nulling signal. The DAN loop also creates a virtual
ground at the $L_\mathrm{SQUID}$ input, dramatically reducing the effective
input impedance of the SQUID inductor.

\subsection{Digital Signal Path}

\begin{figure*}
	\centering
	\begin{tikzpicture}[node distance=3ex, auto]
		\node[dspsquare, text width=3em] (adc) {ADC};
		\node[inner sep=0, minimum size=0, right=of adc,label=${D'[k]}$] (adc_node) {};
		\draw[dspline] (adc) -- (adc_node);

		% chain 1 mixer to mixer
		\node[dspmixer, right=of adc_node] (mix1d) {};
		\node[dspsquare, right=of mix1d, text width=3em] (lpf1d) {LPF};
		\node[dspsquare, right=of lpf1d] (dec1) {$\downarrow 32$};
		\node[inner sep=0, minimum size=0, right=of dec1,label=below:${d_1[l]}$] (d1_node) {};
		\node[dspsquare, right=of d1_node, text width=4em] (dan1) {$\frac{K_1}{1-z^{-1}}$};
		\node[inner sep=0, minimum size=0, right=of dan1,label=below:${n_1[l]}$] (dan1_node) {};

		\node[dspsquare, right=of dan1_node] (int1) {$\uparrow 32$};
		\node[dspsquare, right=of int1, text width=3em] (lpf1u) {LPF};
		\node[dspmixer, right=of lpf1u] (mix1u) {};

		\draw[dspconn] (adc_node) |- (mix1d.west);
		\draw[dspconn] (mix1d.east) |- (lpf1d);
		\draw[dspconn] (lpf1d.east) |- (dec1);
		\draw[dspconn] (dec1.east) |- (dan1);
		\draw[dspline] (dan1.east) |- (dan1_node);
		\draw[dspconn] (dan1_node) |- (int1);
		\draw[dspconn] (int1.east) |- (lpf1u);
		\draw[dspconn] (lpf1u.east) |- (mix1u.west);

		% chain 1 readout
		\node[dspsquare, above=of dan1, text width=5em] (readout1) {Readout};
		\draw[dspconn,dashed] (d1_node) |- (readout1.west) {};
		\draw[dspconn] (dan1_node) |- (readout1.east) {};

		% chain 1 synthesizers
		\node[dspsine, above=of mix1d,dsp/label=right] (dds1d) {$e^{j(-2\pi f_1 k/F_s + \theta^\mathrm{D}_1)}$};
		\draw[dspconn] (dds1d.south) -- (mix1d.north);
		\node[dspsine, above=of mix1u,dsp/label=left] (dds1u) {$e^{j(2\pi f_1 k/F_s + \theta^\mathrm{N}_1)}$};
		\draw[dspconn] (dds1u.south) -- (mix1u.north);

		% Enclosing boxes
		\node[draw,dotted,fit=(mix1d) (lpf1d) (dec1) (dds1d),label=DDC] {};
		\node[draw,dotted,fit=(mix1u) (lpf1u) (int1) (dds1u),label=DUC] {};

		% ellipsis
		\node[below=of mix1d, below=1ex] (ddc_ellipsis) {$\vdots$};
		\node[below=of mix1u, below=1ex] (duc_ellipsis) {$\vdots$};

		% chain 2 mixer to mixer
		\node[dspmixer, below=of ddc_ellipsis] (mix2d) {};
		\node[dspsquare, right=of mix2d, text width=3em] (lpf2d) {LPF};
		\node[dspsquare, right=of lpf2d] (dec2) {$\downarrow 32$};
		\node[inner sep=0, minimum size=0, right=of dec2,label=above:${d_M[l]}$] (d2_node) {};
		\node[dspsquare, right=of d2_node, text width=4em] (dan2) {$\frac{K_M}{1-z^{-1}}$};
		\node[inner sep=0, minimum size=0, right=of dan2, label=${n_M[l]}$] (dan2_node) {};

		\node[dspsquare, right=of dan2_node] (int2) {$\uparrow 32$};
		\node[dspsquare, right=of int2, text width=3em] (lpf2u) {LPF};
		\node[dspmixer, right=of lpf2u] (mix2u) {};

		\draw[dspconn] (adc_node) |- (mix2d.west);
		\draw[dspconn] (mix2d.east) |- (lpf2d);
		\draw[dspconn] (lpf2d.east) |- (dec2);
		\draw[dspconn] (dec2.east) |- (dan2);
		\draw[dspline] (dan2.east) |- (dan2_node);
		\draw[dspconn] (dan2_node) |- (int2);
		\draw[dspconn] (int2.east) |- (lpf2u);
		\draw[dspconn] (lpf2u.east) |- (mix2u.west);

		% chain 2 readout
		\node[dspsquare, below=of dan2, text width=5em] (readout2) {Readout};
		\draw[dspconn,dashed] (d2_node) |- (readout2.west) {};
		\draw[dspconn] (dan2_node) |- (readout2.east) {};

		% chain 2 synthesizers
		\node[dspsine, below=of mix2d,dsp/label=right] (dds2d) {$e^{j(-2\pi f_M k/F_s + \theta^\mathrm{D}_M)}$};
		\draw[dspconn] (dds2d.north) -- (mix2d.south);
		\node[dspsine, below=of mix2u,dsp/label=left] (dds2u) {$e^{j(2\pi f_M k/F_s + \theta^\mathrm{N}_M)}$};
		\draw[dspconn] (dds2u.north) -- (mix2u.south);

		% Enclosing boxes
		\node[draw,dotted,fit=(mix2d) (lpf2d) (dec2) (dds2d)] {};
		\node[draw,dotted,fit=(mix2u) (lpf2u) (int2) (dds2u)] {};

		% summer and DAC
		\node[dspadder, right=of mix1u] (summer) {};
		\draw[dspconn] (mix1u.east) -- (summer.west);
		\draw[dspconn] (mix2u.east) -| (summer.south);

		\node[dspsquare, right=of summer, text width=3em] (real) {$\mathbb{R}\{\cdot\}$};
		\node[inner sep=0, minimum size=0, right=of real, label=${N'[k]}$] (sum_node) {};
		\node[dspsquare, right=of sum_node,text width=3em] (dac) {DAC};

		\draw[dspconn] (summer.east) -- (real);
		\draw[dspline] (real) -- (sum_node);
		\draw[dspconn] (sum_node) -- (dac);

	\end{tikzpicture}

	\caption{A representative model of the readout and nuller signal path
		at RF, neglecting implementation-specific latency imposed by the DUC/DDCs.
		The SQUID output, after
		amplification and digitization by the ADC (left), is presented to an array of
		complex downconverters with distinct frequencies $f_m$ and phases $\theta^\mathrm{D}_m$.
		These signals are decimated to a lower sampling rate (from 20 MSPS to 625
		ksps), where each signal is presented to a distinct integral controller with
		per-channel loop gain $K_m$. Each integrator's output is then
		upconverted back to RF, and the results are summed and presented to the nuller
		DAC (right); see~\Cref{eqn:carrier-sinusoid}.
		Science data may be selected from the open-loop demodulator
		path $d_m[l]$ or closed-loop nuller path $n_m[l]$, and is
		separately decimated and streamed for offline analysis.
	}
	\label{fig:full-signal-path}
\end{figure*}
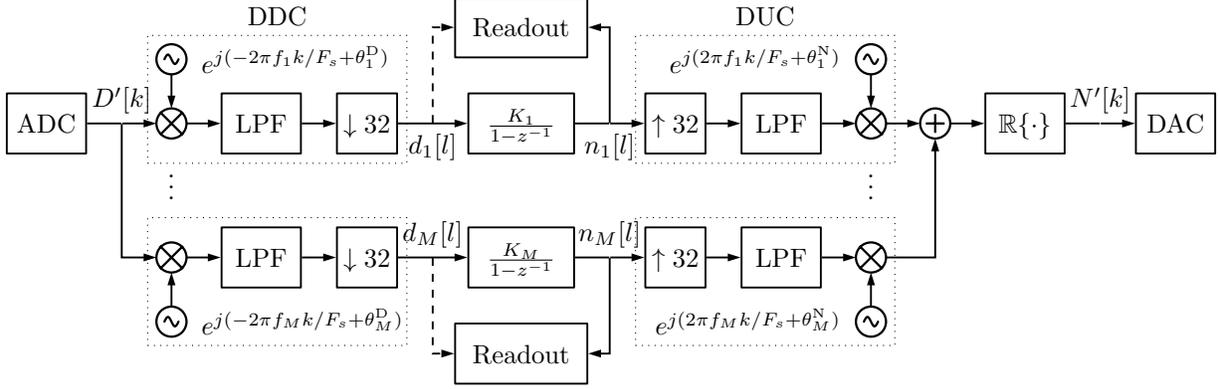

The digital signal path, shown in~\Cref{fig:full-signal-path}, is responsible
for three tasks:

\begin{itemize}
\item Synthesizing carrier and nuller combs for the DACs,
\item Demodulating, decimating, packaging, and streaming bolometer data for analysis, and
\item Using demodulator data to update the synthesizer coefficients used for nulling.
\end{itemize}

The carrier and nuller signals, which are discrete-time versions of $C(t)$ and
$N(t)$, are of the form:

\begin{equation}
	C'[k]=\mathbb{R}\left\{ \sum_{m=1}^M C_me^{2\pi j f_m k/F_s + j \theta^\mathrm{C}_m} \right\}
\label{eqn:carrier-sinusoid}
\end{equation}

\begin{equation}
	N'[k]=\mathbb{R}\left\{ \sum_{m=1}^M n'_m[k]e^{2\pi j f_m k /F_s + j \theta^\mathrm{C}_m} \right\}
\label{eqn:nuller-sinusoid}
\end{equation}

where the signal $n'_m[k]$ is an oversampled version of the baseband nuller
signal $n_m[l]$ shown in~\Cref{fig:full-signal-path}. Since the specific
implementation of the digital up- and down-converters is not the subject of the
present investigation, we leave the exact relationship between the
high-sampling-rate signals $D'[k]$, $n'_m[k]$, and $N'[k]$ and their baseband
counterparts $d_m[l]$ and $n_m[l]$ unspecified here.

Although~\Cref{fig:full-signal-path} does not show the carrier synthesizer path,
it is a degenerate version of the nuller upconverter using programmable but
constant amplitude and phase parameters $C_m$ and $\theta_m^\mathrm{C}$. For the
purposes of evaluating DAN loop performance, the carrier signal is stimulus and
hence not relevant to stability.

In practice, implementing the digital upconverters (DUCs) and digital
downconverters (DDCs) as implied by~\Cref{fig:full-signal-path}
and~\Cref{eqn:carrier-sinusoid,eqn:nuller-sinusoid} is not computationally
efficient (and hence, not power- or resource-efficient)~\cite{Smecher2012}.
Rather than implement independent up-/down-converters for each bolometer, we
use intermediate up- and down-conversion stages using polyphase filter-bank
(PFB) techniques~\cite{harris2004}.  We also do not synthesize (and then
discard) the imaginary portion of the carrier and nuller signals. In exchange
for a more efficient implementation, the use of PFB techniques imposes
relatively high latency on the up- and downconversion chains.

Signal path designs to date have exhibited combined nuller and demodulator
latencies between \SI{5}{\micro \second} and \SI{31}{\micro \second}; the
current signal path design has a combined nuller and demodulator latency of
approximately $\tau \approx \SI{14}{\micro \second}$.

The specific implementation of the DDC and DUC blocks are not the subject of
the present investigation; here, only the latency term $\tau$ is important.
We will return to the latency term after describing how such a system is tuned
in practice.

\subsection{Tuning and Operation}

The signal path in~\Cref{fig:full-signal-path} and the signals
in~\Cref{eqn:carrier-sinusoid,eqn:nuller-sinusoid} have many parameters
(carrier gains, phases, loop gains, and frequencies). To motivate how and when
these parameters are selected, a simplified description of a typical array
tuning algorithm follows.

\begin{enumerate}
	\item
		The array starts off at a temperature above the bolometers'
		superconducting transition $T_c$ (typically \SI{4}{\kelvin}, the
		temperature of the cryostat's liquid Helium bath); hence,
		bolometers exhibit a nominal resistance regardless of any
		optical power absorbed or electrical power dissipated.
		Signal-path frequencies $f_m$ are programmed using a database
		of pre-characterized parameters. Amplitude settings ($C_m,
		K_m$) ensure $C(t)=N(t)=0$. Data readout uses the open-loop
		(dashed) path in~\Cref{fig:full-signal-path}.
	\item
		The readout phases $\theta^\mathrm{D}_m$ are tuned to rotate
		each bolometer's nuller signal into the in-phase or I component
		of the corresponding complex readout signal $d_m[l]$, with an
		\SI{180}{\degree} phase shift to ensure negative feedback.
	\item
		The DAN loop can now be closed. To do this, nonzero gain terms
		$K_m$ are programmed. Nullers begin to respond dynamically.
		Capture of science data is shifted from open-loop paths (the
		dashed lines in~\Cref{fig:full-signal-path}) to closed-loop
		paths (solid lines).
	\item
		Carrier amplitudes $C_m$ are programmed, delivering enough
		electrical power to each bolometer to prevent it from latching
		in its superconducting state when the array is cooled below
		$T_c$. Specific $C_m$ values are taken from a database of
		pre-characterized parameters.
	\item
		The array is now cooled below the $T_c$ of the bolometers using
		a sub-Kelvin refrigerator.  Bolometers are now held in
		electrothermal feedback~\cite{Irwin1995}; the current flowing
		through them is a strong function of incident optical power.
	\item
		Carrier phases $\theta^\mathrm{C}_m$ are now tuned into the I
		component of the readout signal.  Since the sidebands generated
		by optical modulation are symmetric, the science signal is
		entirely present in I; the Q channel may be kept for diagnostic
		purposes or discarded.
\end{enumerate}

At this point, the detector array and readout signal path are ready to capture
science data.  However, the tuning process required values for the $M$ DAN
gains $K_m$, and we have not provided a physical or conceptual basis for the
choice of these parameters.

In the remainder of this paper we establish the criteria for choosing a
suitable DAN gain that is both bounded-input-bounded-output (BIBO) stable and
effective at nulling the SQUID input.

\section{The Simplified DAN Loop}

A complete model of the feedback loop and surrounding system would include both
the analog system described in \Cref{fig:cold-schematic} and the digital system
described in \Cref{fig:full-signal-path}. Such a model would be comprehensive,
but would prevent a convenient stability analysis.  To find a tractable model
of the DAN loop, we begin with some simplifying assumptions.

Since the LC combs (and indeed, the bolometers themselves) are part of the
carrier path and do not form part of the feedback loop, they can be removed
when considering loop stability and performance. To do so, we consider the
feedback loop's input to be current injected into the inductance
$L_\mathrm{SQUID}$.\footnote{In reality, the impedances of the carrier RLC networks are
non-negligible relative to the SQUID input impedance, so some nuller current
intended for the SQUID coil inevitably flows back into the associated
bolometer. Although this parasitic effect is critical when designing the cold
electronics, it requires a separate treatment and does not impact the selection
of $K_m$; hence, we neglect it for the present discussion.}

We assume that DAN loops are adequately spaced in frequency and do not
interact. This permits us to decompose the system into $M$ individual DAN
loops, and analyze the stability and performance of just one.

We assume that all RF components in the system are wideband relative to the
bandwidth of the digital loop. (In current designs, operates at a
sampling rate of \SI{625}{\kilo \hertz} and has meaningful gain only at a small
fraction of this bandwidth.) These frequency-dependent terms are replaced with
constant gains.

We recognize that the analog system is now frequency-independent. Gains can be
freely moved across the downconverter to baseband. The DAC and ADC cancel each
other out (any time delays removed by this cancellation are minor but can be
re-injected using the latency model added next), followed by cancellation of the
upconverter and downconverter under the assumption that the phase terms
$\theta_m^\mathrm{N}$ and $\theta_m^\mathrm{D}$ are aligned during tuning as
described above. The entire model is now discrete-time and operates at
baseband.

Finally, we need to inject a latency model. Because the system is now fully
discrete-time, It makes sense to do this in the z-domain where delays in
integer sample counts are convenient.  We add a delay of $L$ samples in the
feedback path. In our system, a latency of $\tau \approx \SI{14}{\micro \second}$ is
equivalent to $L=9$ at a baseband sampling rate of \SI{625}{\kilo \hertz}.

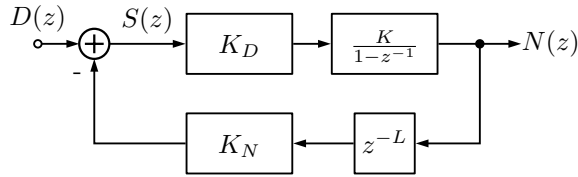
\begin{figure}[H]
	\centering
	\begin{tikzpicture}[node distance = 0.5cm, auto]
		\node[dspnodeopen] (input) {$D(z)$};
		\node[dspadder,right=of input,dsp/label=above] (summer) {};
		\node[inner sep=0, minimum size=0, right=of summer, label=above:$S(z)$] (summer_node) {};
		\node[inner sep=0, minimum size=0, below left=3ex and 0.5ex of summer, label=-] (sum_annotation) {};
		\node[dspsquare,right=of summer_node, text width=4em] (hd) {$K_D$};
		\node[dspsquare,right=of hd, text width=4em] (dan) {$\frac{K}{1-z^{-1}}$};
		\node[dspnodefull, right=of dan] (turnaround) {};
		\node[inner sep=0, minimum size=0, right=of turnaround] (output) {$N(z)$};

		\node[dspsquare,below=of dan] (latency) {$z^{-L}$};
		\node[dspsquare,below=of hd, text width=4em] (hn) {$K_N$};

		\draw[dspconn] (input) -- (summer.west);
		\draw[dspconn] (summer.east) -- (hd);
		\draw[dspconn] (hd) -- (dan);
		\draw[dspline] (dan) -- (turnaround);
		\draw[dspconn] (turnaround) |- (latency);
		\draw[dspconn] (latency.west) |- (hn);
		\draw[dspconn] (hn) -| (summer.south);
		\draw[dspconn] (turnaround) -- (output);

	\end{tikzpicture}

	\caption{A model of the DAN feedback loop at baseband}
	\label{fig:simpler-dan-loop}
\end{figure}

After these simplifications, a single DAN loop may be modelled as shown
in~\Cref{fig:simpler-dan-loop}.

This model includes the SQUID (summing junction), gains for the nuller ($K_N$)
and demodulator ($K_D$) circuitry, the DAN integrator transfer function, and
the latency term $z^{-L}$.

The closed-loop transfer function of this system at baseband is

\begin{equation}
	H(z) = \frac{N(z)}{D(z)} = \frac{K_DK}{1-z^{-1} + K_DK_NKz^{-L}} = \frac{K_DK}{1-z^{-1} + K_0z^{-L}}
\label{eqn:h}
\end{equation}

The transfer function from baseband input to the residual (nulled) signal seen
by the SQUID is

\begin{equation}
	E(z) = \frac{S(z)}{D(z)} = \frac{1-z^{-1}}{1-z^{-1}+K_NK_DKz^{-L}} = \frac{1-z^{-1}}{1-z^{-1}+K_0z^{-L}}
\label{eqn:e}
\end{equation}

where in both equations we have defined the combined gain $K_0=K_NK_DK$.

\subsection{Stability}

Our first consideration is given to stability. This system is stable when its
z-plane poles are within the unit circle; hence, we are interested in the
impact of gain $K_0$ and latency $L$ on the poles of $H(z)$.

Both the closed-loop and residual transfer functions have the same
characteristic equation (i.e. where the denominator of the transfer function
equals 0):

\begin{equation}
z^L-z^{L-1}+K_0=0
\end{equation}

As this equation is not soluble in closed form for typical $L$, we resort to
numerical techniques. We show some results for stability versus latency and
loop gain later (see~\Cref{fig:resonance-vs-kl}). As expected, loop stability
diminishes with increasing $K_0$ and $L$.

For the remainder of this paper we are only interested in the performance of
stable loop configurations. We now consider nulling performance within the
stable regime.

\subsection{Nulling Performance}

\begin{figure}
	\centering
	\includegraphics{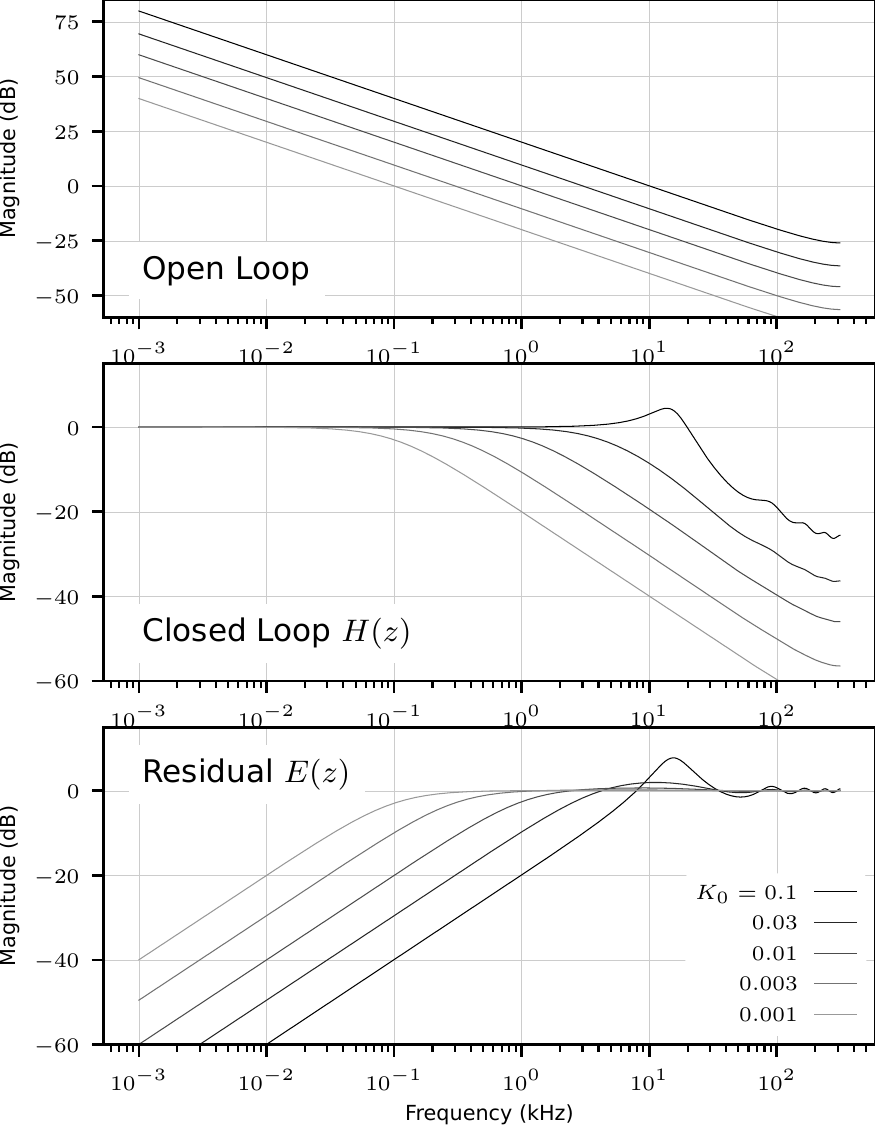}
	\caption{System transfer function vs. frequency, with constant latency $L=9$ and
		varying gain $K_0$. The gain is varied logarithmically from $K_0=10^{-3}$
		to $K_0=10^{-1}$.}
	\label{fig:tf-vs-gain}
\end{figure}

\Cref{fig:tf-vs-gain} shows the open-loop magnitude, closed-loop magnitude
$H(z)$, and residual magnitude $E(z)$ versus frequency as the gain $K_0$ is
varied.  All gains in this range are stable for $L=9$. However, at higher $K_0$,
there are regions where the magnitude response of the nuller loop is greater
than \SI{0}{\deci \bel} (that is, $|E(e^{j\omega})|^2 > 1$). Though stable,
loop gains with this characteristic are not suitable for nulling because they
amplify rather than suppress some inputs at the SQUID.

\begin{figure}
	\centering
	\includegraphics{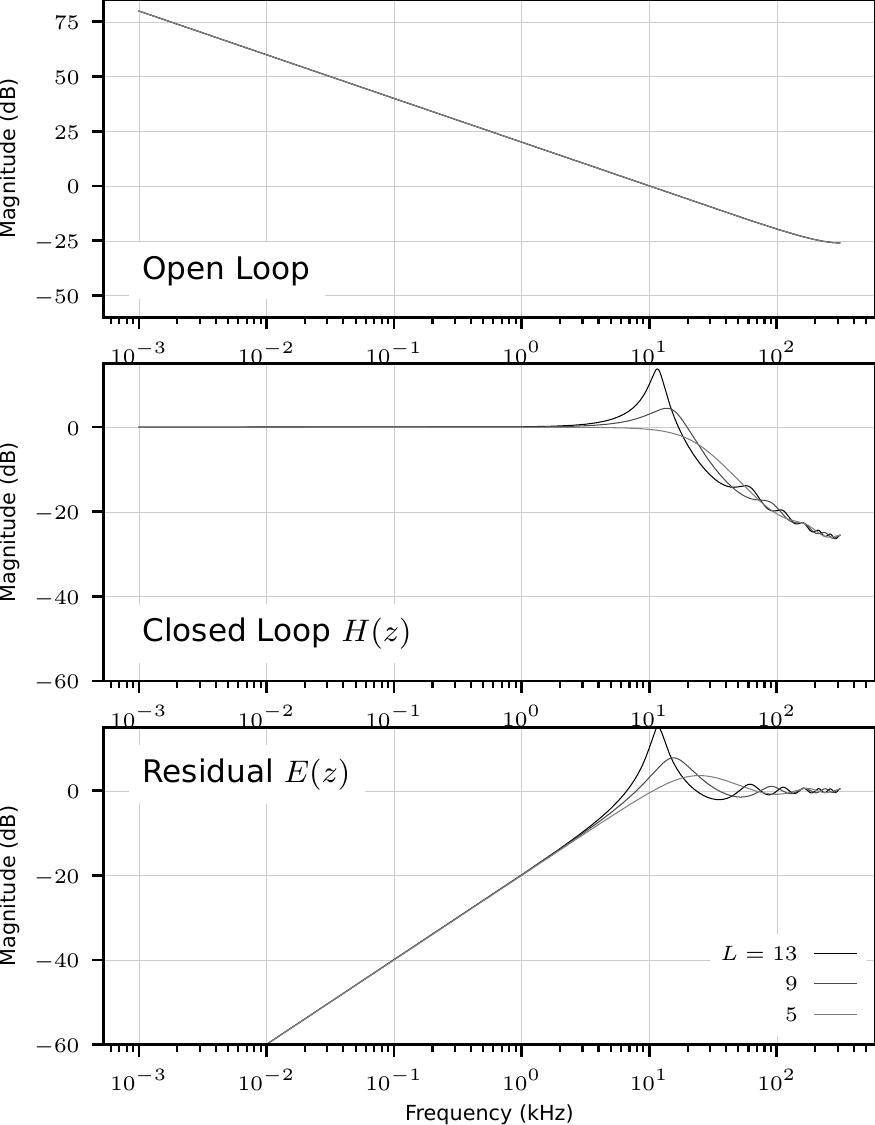}
	\caption{System transfer function vs. frequency, with constant gain
		$K_0=0.1$ and varying latency $L$.}
	\label{fig:tf-vs-latency}
\end{figure}

Similarly,~\Cref{fig:tf-vs-latency} shows the transfer functions $H(z)$ and
$E(z)$ for varying latency $L$ under a constant gain $K_0=0.1$. Unlike loop gain
$K_0$, latency is essentially a design parameter; by varying $L$ we are
exploring the performance available to a system with that latency.
Stable designs with sufficiently high $L$ also begin to exhibit anti-nulling
behaviour.

Correct interpretation of these peaks is crucial.  Let us suppose we are
operating a DAN loop at a sufficient gain $K_0$ that the peak in these plots is
present.  Since this is a stable feedback loop, this system (both $H(z)$ and
$E(z)$) does not \emph{create} signals at these frequencies -- but any signals
already present within this region are amplified before being presented to the
SQUID (in the case of $E(z)$) or synthesized by the nuller (in the case of
$H(z)$). In practice, a signal in this region could be injected into the DAN
loop from several sources:

\begin{itemize}
	\item
                The signal could come from a predictable noise feature (e.g.
                from amplifiers in the carrier, nuller, or demodulator chain,
                or from the noise floor of the DACs, ADCs, or digital signal
                chain).
	\item
		The signal could come from a neighbouring bolometer. (This is
		unlikely, since in practice bolometer LCs are spaced much
		further apart than the spectral peaks at approximately
		\SI{10}{\kilo\hertz} in these design examples.)
	\item
		The signal could come from EMI, an aliasing image,
		intermodulation distortion, or some other spurious narrowband
		signal. By themselves, this kind of signal consumes SQUID
		dynamic range (contributing to nonlinear distortion). When
		amplified by an incorrectly configured DAN loop, the aggressor
		signal consumes even more of the SQUID dynamic range, and the
		DAN loop consumes dynamic range from the nuller DAC and signal
		chain to do so.
\end{itemize}

Although we could operate a DAN loop in a region exhibiting peaks, it leaves
the feedback loop susceptible to noise sources that are poorly characterized
and difficult to control. In a system with many thousands of bolometers at
frequencies determined by LC combs that are known to scatter during cooldown,
there are likely to be coincidences. Better, then, to choose a loop gain at
which this amplification effect is known not to occur.

\begin{figure}
	\centering
	\includegraphics{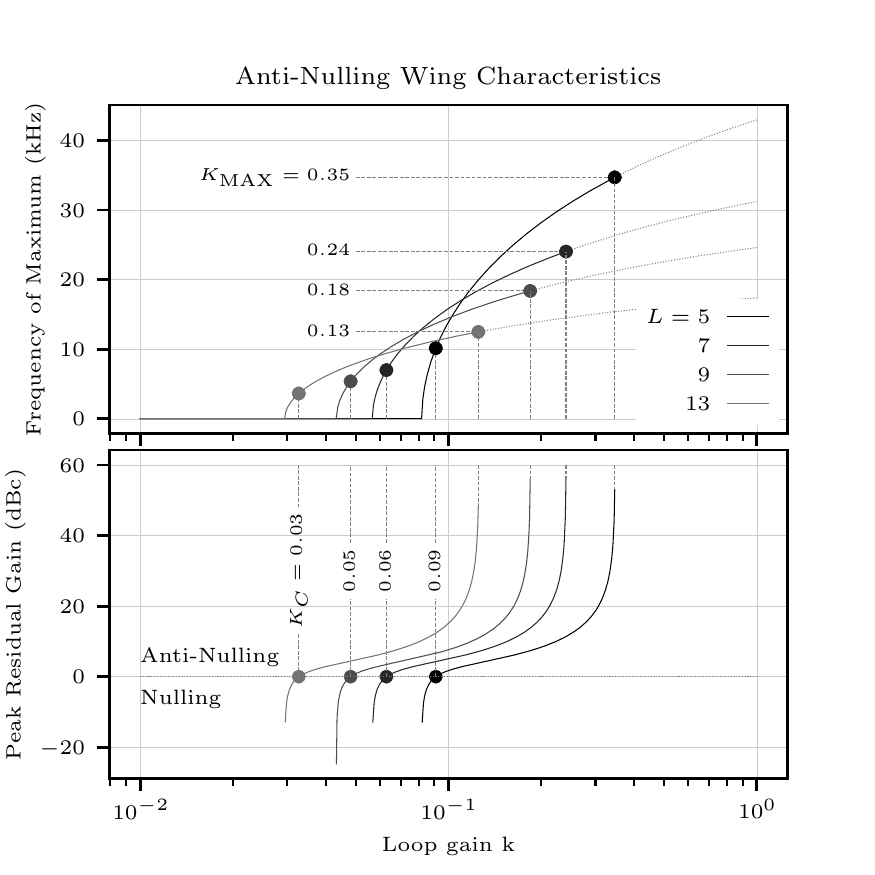}
	\caption{The frequency and gain of spectral peaks in the residual transfer function $E(z)$,
		as a function of loop gain $K_0$. The full range of stable loop
		gains $K_0$ is shown; when the loop becomes unstable (i.e. when
		the poles move outside the unit circle) the line is shown
		dashed. The maximum BIBO stable gain is marked $K_\mathrm{MAX}$
		for each latency. We also show the critical loop gain $K_C$ for which the
		wing amplitude (i.e. amplification factor) crosses unity.
		For effective nulling, the loop gain should never cross
		this critical gain $K_C$.
	}
	\label{fig:resonance-vs-kl}
\end{figure}

\Cref{fig:resonance-vs-kl} shows the magnitude and frequency of spectral peaks
as a function of loop gain $K_0$ for latencies $L=5$, $L=7$, $L=9$, and $L=13$.
For each of these latency scenarios, there exist two critical gains $K_0$:

\begin{itemize}
	\item
		$K_\mathrm{MAX}$, beyond which the loop is no longer BIBO
		stable, and
	\item
		$K_C$, the ``critical gain'' at which anti-nulling
		(amplification) starts to occur.
\end{itemize}

The ratio $K_\mathrm{MAX}/K_C$ is remarkably independent of latency, and is
approximately 3.8 for all latencies between $L=2$ and $L=20$.

\section{Experimentally Deriving DAN Gain}

Recall that the combined gain $K_0=K_NK_DK$, for which we have been evaluating
performance, is the product of the nuller gain $K_N$, demodulator gain $K_D$,
and digital gain $K$. The nuller and demodulator gains $K_N$ and $K_D$ are
themselves products of many gain terms, including terms which are easy to
control (amplifier gains, DAC/ADC conversion ratios) and those that are
inconvenient to precisely control (cryogenic components, analog filter
roll-offs). The uncertainties in these terms compound, leaving an overall
system with a great deal of gain variation from bolometer to bolometer on a
single SQUID, and from SQUID to SQUID in a larger system.

Although most of these gain terms are fabrication variables and can be
measured, the practical complications associated with measuring, tracking, and
correctly compensating for these gains are undesirable in an experiment with
many thousands of bolometers.

In the past, we accommodated scatter in $K_N$ and $K_D$ by using a digital
gain $K$ in the middle of a relatively wide ``Goldilocks'' region, where
nulling efficacy is adequate (even when $K_NK_D$ scatters unexpectedly low) and
amplification effects and instabilities have not begun to occur (even when
$K_NK_D$ scatters high). As latency increases, however, the size of this
Goldilocks region shrinks and we either need to characterize the system better
or use another approach.

During the nuller alignment step in the tuning algorithm described above, we
injected a fixed nuller signal into the system in order to rotate the nuller
phase $\theta_m^\mathrm{N}$ into anti-alignment with the demodulator (i.e. to correctly
establish negative feedback). However, the \emph{magnitude} of the nuller
injection gives us a measurement of $K_NK_D$.  Hence, we developed the
following procedure to establish a consistent loop gain $K_0 < K_C$:

\begin{itemize}
	\item
		Choose a desired gain $K_0 < K_C$, e.g.
		using~\Cref{fig:resonance-vs-kl}.
	\item
		During nuller phase alignment, as described above, a small
		nuller signal is injected. This causes a complex displacement
		in the readout signal.
	\item
		The angle of this displacement, as described above, is used to
		determine the nuller phase $\theta_m^\mathrm{N}$ in order to establish
		negative feedback.
	\item
		The magnitude of this displacement, divided by the amplitude of
		the nuller probe tone, is a direct measurement of the gain
		$K_NK_D$.  The digital gain parameter $K$ should be set to
		$K=K_0/K_NK_D$.
\end{itemize}

This procedure does not require any additional data capture during tuning, and
improves consistency across the system even in systems with low enough latency
that detailed gain characterization is unnecessary.

\section{Conclusions}

In this paper, we evaluated the effects of latency $\tau$ (equivalently, L) on
a digital active nulling (DAN) feedback loop used in frequency-multiplexed
bolometer readout. We characterized the impact of latency on both BIBO loop
stability and the ability to effectively null an input signal.

We found that BIBO-stable gains $K_0$ between the loop's maximum stable gain
$K_\mathrm{MAX}$ and some critical gain $K_C\approx K_\mathrm{MAX}/3.8$ were
susceptible to spurious signals within the loop bandwidth, which would be
amplified (rather than suppressed) at the SQUID input coil. The frequency of
problematic signals depended on both $L$ and $K_0$.  Although a DAN system can
operate above $K_C$ in the absence of aggressor signals, we recommend staying a
safe margin below $K_C$ to avoid unpredictable effects in high-bolometer-count
systems where spurious signals are not always well characterized, predictable,
or controllable. %This recommendation matches the heuristic choice $K_0 =
%K_\mathrm{MAX}/3$ that has been successfully used for some time.

Finally, we proposed an improved mechanism for determining $K$ to achieve a
desired overall loop gain $K_0$ in the presence of poorly characterized gains
$K_N, K_D$. This tuning step takes advantage of measurements that are already
required when establishing phase-aligned negative feedback in the nuller. This
procedure compensates for channel-to-channel variability in loop gain due to
analog parameter variations (and hence, variability in nulling performance).
Loop-gain compensation is desirable even when latency ($L, \tau$) is small
enough that the loop performs suitably over a wide enough range of $K$ to
accommodate variations in $K_N$ and $K_D$.

\section*{Acknowledgements}

The McGill authors acknowledge funding from the Natural Sciences and
Engineering Research Council of Canada (NSERC), Canadian Institute for Advanced
Research (CIFAR), and Fonds de Recherche Nature et Technologies (FRQNT).

Tijmen de Haan acknowledges the LBNL Chamberlain Fellowship and funding from the
Quantum Sensors HEP-QIS Consortium and the LBNL Laboratory Directed Research
and Development (LDRD) program under U.S. Department of Energy Contract No.
DE-AC02-05CH11231.

\bibliography{dan_memo.bib}{}
\bibliographystyle{spiebib}

\end{document}